# COMPLICATED BACKGROUND SUPPRESSION OF VISAR IMAGE FOR MOVING TARGET SHADOW DETECTION

*Zhenyu Yang, Xiaoling Zhang, Xu Zhan*

School of Information and Communication Engineering,
University of Electronic Science and Technology of China, Chengdu 611731, P.R.China

**ABSTRACT**

The existing Video Synthetic Aperture Radar (ViSAR) moving target shadow detection methods based on deep neural networks mostly generate numerous false alarms and missing detections, because of the foreground-background indistinguishability. To solve this problem, we propose a method to suppress complicated background of ViSAR for moving target detection. In this work, the proposed method is used to suppress background; then, we use several target detection networks to detect the moving target shadows. The experimental result shows that the proposed method can effectively suppress the interference of complicated background information and improve the accuracy of moving target shadow detection in ViSAR.

## 1. INTRODUCTION

Video Synthetic Aperture Radar (ViSAR) technology is an extension of the existing SAR system. This technology combines the advantages of high-resolution imaging and video imaging. ViSAR records the changes in the target area, and presents the information in the form of visual active images conducive to the intuitive interpretation in the time dimension through signal processing. It can obtain the important information such as high-precision position and speed of the moving target in real time, and carry out continuous tracking and monitoring.

Detecting and tracking moving targets is the main task of ViSAR, and moving target shadow detection is the key technique to complete this task. With the wide application of deep learning in various fields, some researchers also apply it in ViSAR moving target detection. Ding et al. [1] used the improved faster region-based convolutional neural network (Faster-RCNN) for moving target shadow detection. Zhang et al. [2] proposed a moving target tracking method based on shadow detecting and tracking with convolution neural network (CNN). Zhou et al. [3] presented a deep network using CNN and long short-term memory (LSTM) module to detect and track the shadow of ground moving target in ViSAR.

For moving target shadow detection of ViSAR, false alarm often appears in complicated background. At the same time, moving target detection relies on shadows, which are often difficult to distinguish from low-scattering areas such as roads, which can cause a lot of misses. However, the methods mentioned above all focus on improving the networks for shadow detecting and tracking. Differently, for improving foreground-background distinction, a complicated background suppression method based on decomposition of background, target and noise is proposed, and resolved by alternating direction method of multipliers (ADMM). The experimental result shows the effectiveness of our method for moving target shadow detection in ViSAR.

The rest of this paper is organized as follows: Section II proposes our method. Section III provides the experimental result and analysis. Conclusions is presented in Section IV.

## 2. METHOD

In this section, the background suppression method is presented. When we stack each frame as a column of a matrix, the video can be reshaped as a two-dimensional matrix $\mathbf{D}$. Most of the background information is similar, it contains a lot of redundant information, which can be represented by a low-rank matrix $\mathbf{B}$, and the foreground with moving targets contains few non-zero elements, so the foreground matrix $\mathbf{X}$ is sparse. Considering that most of the noise obeys a Gaussian distribution, matrix $\mathbf{E}$ denotes the white Gaussian noise. Then $\mathbf{D}$ can be expressed as $\mathbf{D} = \mathbf{B} + \mathbf{X} + \mathbf{E}$.

In order to decompose the matrices of background, target and noise, it can be transformed as the following optimization problem [4]:

$$\min_{\mathbf{B},\mathbf{X},\mathbf{E}} rank(\mathbf{B}) + \xi \|\mathbf{X}\|_0 + \gamma \|\mathbf{E}\|_F^2 \qquad (1)$$
$$s.t. \mathbf{D} = \mathbf{B} + \mathbf{X} + \mathbf{E}$$

where $\mathbf{D}, \mathbf{B}, \mathbf{X}, \mathbf{E} \in \mathbb{R}^{m \times n}$ ($m$ is the product of the height and width of the frame, and $n$ is the number of the frame), $rank(\cdot)$ denotes the rank of matrix, $\|\cdot\|_0$ denotes the $\ell_0$ norm, and $\|\cdot\|_F$ denotes the Frobenius norm.

Both $rank(\mathbf{B})$ and $\|\mathbf{X}\|_0$ contained in the objective function are non-convex nonlinear combinatorial optimization functions, so its solution is an NP-Hard problem.

Hence, through augmented Lagrange multipliers optimization, the foreground can be reconstructed as:

$$L(\mathbf{B},\mathbf{X},\mathbf{E},\mathbf{Y}) = \|\mathbf{B}\|_* + \xi\|\mathbf{X}\|_1 + \gamma\|\mathbf{E}\|_F^2 \\ + \frac{\mu}{2}\|\mathbf{D}-\mathbf{B}-\mathbf{X}-\mathbf{E}\|_F^2 + \langle\mathbf{Y},\mathbf{D}-\mathbf{B}-\mathbf{X}-\mathbf{E}\rangle \quad (2)$$

where $\|\cdot\|_*$ denotes the nuclear norm and $\|\cdot\|_1$ denotes the $\ell_1$ norm, $\xi$ and $\gamma$ denotes the soft threshold. $\mathbf{Y}$ denotes a Lagrange multiplier matrix. The objective function is a composite of three terms, the background, target and noise respectively. And the purpose is to suppress the background through three-term decomposition.

We can solve the problem by ADMM [5].

1) Updating $\mathbf{B}$:

$$\mathbf{B}_{k+1} = \arg\min_{\mathbf{B}} L(\mathbf{B},\mathbf{X}_k,\mathbf{E}_k,\mathbf{Y}_k,\mu_k) \\ = \arg\min_{\mathbf{B}} \|\mathbf{B}\|_* + \frac{\mu_k}{2}\left\|\mathbf{B}-\left(\mathbf{D}-\mathbf{X}_k-\mathbf{E}_k+\frac{\mathbf{Y}_k}{\mu_k}\right)\right\|_F^2 \quad (3) \\ = M_{1/\mu_k}\left(\mathbf{D}-\mathbf{X}_k-\mathbf{E}_k+\frac{\mathbf{Y}_k}{\mu_k}\right)$$

2) Updating $\mathbf{X}$:

$$\mathbf{X}_{k+1} = \arg\min_{\mathbf{X}} L(\mathbf{B}_{k+1},\mathbf{X},\mathbf{E}_k,\mathbf{Y}_k,\mu_k) \\ = \arg\min_{\mathbf{X}} \xi\|\mathbf{X}\|_1 + \frac{\mu_k}{2}\left\|\mathbf{X}-\left(\mathbf{D}-\mathbf{B}_{k+1}-\mathbf{E}_k+\frac{\mathbf{Y}_k}{\mu_k}\right)\right\|_F^2 \quad (4) \\ = N_{\xi/\mu_k}\left(\mathbf{D}-\mathbf{B}_{k+1}-\mathbf{E}_k+\frac{\mathbf{Y}_k}{\mu_k}\right)$$

3) Updating $\mathbf{E}$:

$$\mathbf{E}_{k+1} = \arg\min_{\mathbf{E}} L(\mathbf{B}_{k+1},\mathbf{X}_{k+1},\mathbf{E},\mathbf{Y}_k,\mu_k) \\ = \arg\min_{\mathbf{E}} \gamma\|\mathbf{E}\|_F^2 + \frac{\mu_k}{2}\left\|\mathbf{E}-\left(\mathbf{D}-\mathbf{B}_{k+1}-\mathbf{X}_{k+1}+\frac{\mathbf{Y}_k}{\mu_k}\right)\right\|_F^2 \quad (5) \\ = \left(1+\frac{2\gamma}{\mu_k}\right)^{-1}\left(\mathbf{D}-\mathbf{B}_{k+1}-\mathbf{X}_{k+1}+\frac{\mathbf{Y}_k}{\mu_k}\right)$$

4) Updating $\mathbf{Y}$ and $\mu$:

$$\mathbf{Y}_{k+1} = \mathbf{Y}_k + \mu_k(\mathbf{D}-\mathbf{B}_{k+1}-\mathbf{X}_{k+1}-\mathbf{E}_{k+1}) \quad (6)$$

$$\mu_{k+1} = \rho\mu_k \quad (7)$$

where $k(n=1,2\cdots)$ is the label of iteration number. $M_\varepsilon(\mathbf{Q})$ and $N_\varepsilon(\mathbf{Q})$ denotes proximal mapping operators of matrix $\mathbf{Q}$, which can be expressed as:

$$M_\varepsilon(\mathbf{Q}) = UN_\varepsilon(\mathbf{Q})V^\mathrm{T} \quad (8)$$

$$N_\varepsilon(\mathbf{Q})_{ij} = \max(|q_{ij}|-\varepsilon,0)\,\mathrm{sign}(q_{ij}) \quad (9)$$

where $U\Sigma V^\mathrm{T}$ denotes the singular value decomposition of a matrix.

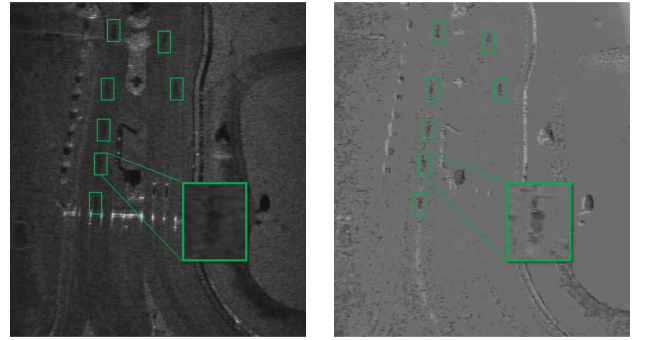

(a)      (b)

**Figure 1.** Results of background suppression at frame 8. Different moving target shadows are marked in green boxes. (a) before background suppression; (b) after background suppression.

**Table 1.** Information entropy result

| Our method | IE |
|---|---|
| ✗ | 8.8025 |
| ✓ | **4.9927 (3.8098↓)** |

**Table 2.** Shadow contrast result

| Our method | Shadow Contrast |
|---|---|
| ✗ | 21.0104 |
| ✓ | **69.3711(48.3607↑)** |

## 3. EXPERIMENTAL RESULT

In this section, the proposed method can be verified by the measured ViSAR data published by Sandia laboratory [6].

### 3.1. Background suppression

First, we use proposed method to suppress background. Figure 1 shows the results of background suppression. After the ViSAR image processed, the background is suppressed distinctly, while the shadows of the moving target are enhanced. Through human eye observation, it can be found that the shadow has become much easier to distinguish. For quantitative evaluation, we use information entropy (IE) to show the background suppression effect. As shown in Table 1, the IE with our method is much lower than without our method (4.9927 << 8.8025), which indicates that the amount of information in the picture (i.e. background) has decreased.

We also evaluate the shadow contrast in the zoom region by using the classic 4-neighborhood method. The shadow contrast results are shown in Table 2. From Table 2, the shadow contrast with our method is far larger than without our method (69.3711 >> 21.0104), and the enhancement reaches up to 230.18%. Hence, through our method, foreground-background distinction is improved significantly.

## 3.2. Detection result

In order to better evaluate the performance of the method proposed in this paper for moving target detection in ViSAR images, the classical target detection networks are used to verify it.

Since a total of 899 SAR images are extracted, 600 frames are used as the train set, and the other 299 consecutive frames are used as the test set. The input image size is 660 × 720. We train networks based on the stochastic gradient descent (SGD) optimizer. We set the learning rate to 0.008 for Faster-RCNN [7] and RetinaNet [8], 0.001 for YOLOv3 [9]. We set the training batch size to 4.

We adopt the evaluation metric of PASCAL VOC, which has 6 indicators to evaluate the detection results, namely $TP$, $FP$, $FN$, Recall, Precision and Average Precision. $TP$ denotes the true positive (correct detections), $FP$ denotes the false positive (false alarms), and $FN$ denotes the false negative (missing detections).

1) Recall ($r$) defined as:
$$r = \frac{TP}{TP+FN} \times 100\% \quad (10)$$

2) Precision ($p$) defined as:
$$p = \frac{TP}{TP+FP} \times 100\% \quad (11)$$

3) Average Precision ($ap$) defined as:
$$ap = \int_0^1 p(r) \cdot dr \quad (12)$$

where $p(r)$ denotes the precision-recall curve. So $ap$ is a comprehensive indicator to evaluate the detection results.

Table 3, given in next page, shows the quantitative results on ViSAR data, where $GT$ denotes the number of ground truths. Compared with Faster-RCNN, after background suppressing, all indicators show the detecting performance improvements. Especially $ap$ is far superior to Faster-RCNN, i.e., 68.43% >> 60.05%, getting an ~8% accuracy improvement. And compared with RetinaNet, $ap$ gets a ~8% accuracy improvement. Also, compared with YOLOv3, $ap$ gets a ~6% accuracy improvement. At the same time, $FN$ and $FP$ have a great reduction, while $TP$, $r$, $p$ and $ap$ are increasing. So complicated background suppression reveals its effectiveness for improve the accuracy of moving target shadow detection, reducing not only the false alarms, but also the missing detections.

Figure 2 shows the detection results of ViSAR images before and after background suppression. It is not difficult to see that background suppression has a significant inhibitory effect on false alarm and missing detection.

## 4. CONCLUSION

This paper proposed a complicated background suppression method for ViSAR moving target shadow detection. Background, target and noise decomposition is used to suppress background to improve foreground-background distinction, which can reduce information entropy and enhance the shadow contrast to reduce false alarms and missing detections. The experimental result verifies the effectiveness of proposed method, and shows significant improvement in moving target shadow detection.

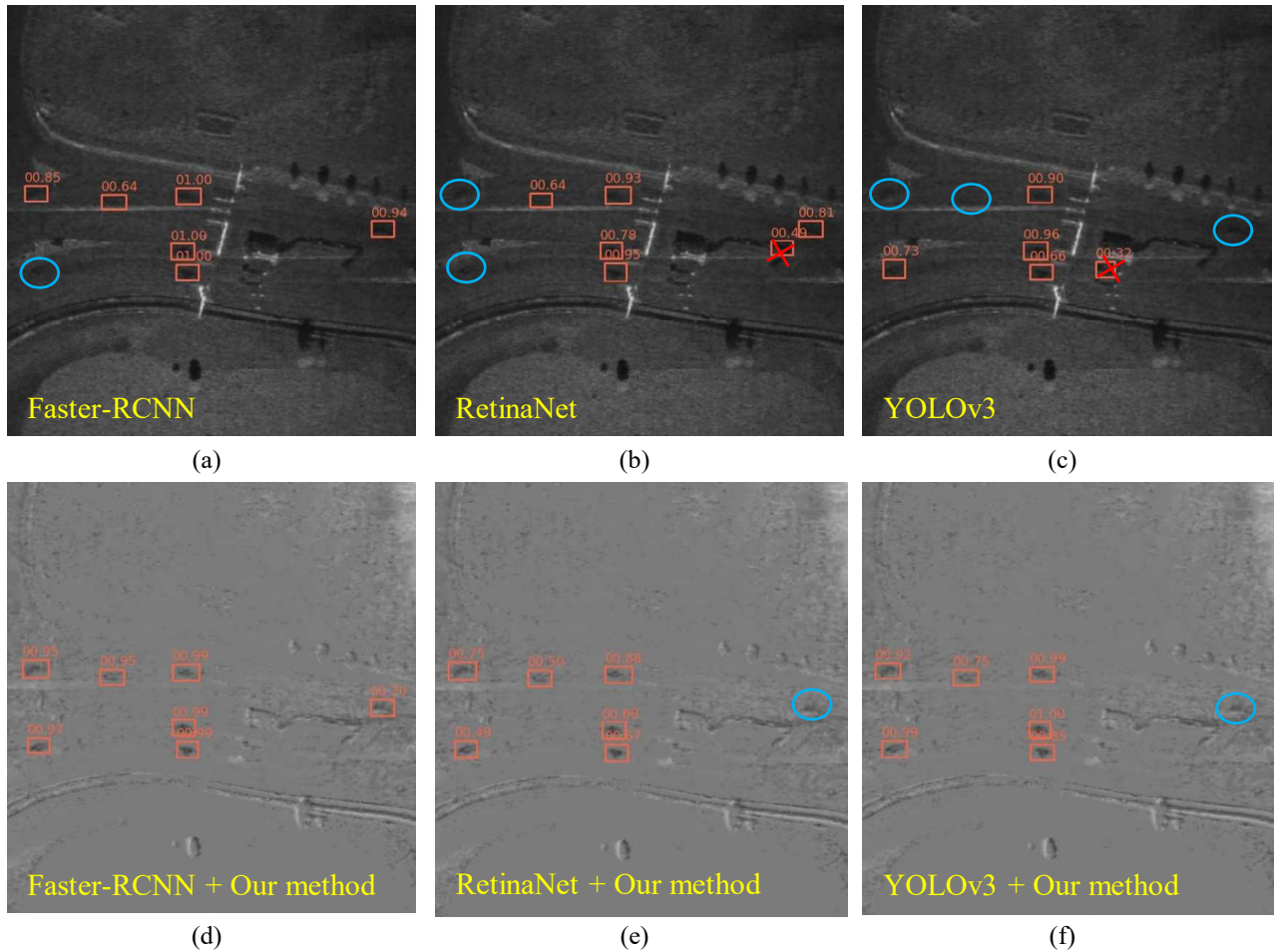

**Figure 2.** ViSAR moving target shadow detection results of different Networks at frame 791. (a) Faster-RCNN; (b) RetinaNet; (c) YOLOv3; (d) Faster-RCNN + Our method; (e) RetinaNet + Our method; (f) YOLOv3 + Our method. The detection results are marked by orange boxes. The false alarms are marked by red crosses. The missing detections are marked by blue ellipses.

Table 3. Quantitative shadow detection results on ViSAR data

| Method | GT | TP | FP | FN | r(%) | p(%) | ap(%) |
|---|---|---|---|---|---|---|---|
| Faster-RCNN | 1702 | 1167 | 584 | 535 | 68.57 | 66.65 | 60.05 |
| Faster-RCNN + **Our method** | 1702 | 1257 | 494 | 445 | 73.85 | 71.79 | **68.43(8.38↑)** |
| RetinaNet | 1702 | 1016 | 580 | 686 | 59.69 | 63.66 | 48.96 |
| RetinaNet + **Our method** | 1702 | 1080 | 401 | 622 | 63.45 | 72.92 | **57.42(8.46↑)** |
| YOLOv3 | 1702 | 965 | 438 | 737 | 56.70 | 68.78 | 49.75 |
| YOLOv3 + **Our method** | 1702 | 1038 | 449 | 664 | 60.99 | 69.80 | **56.34(6.59↑)** |